\def\be#1#2{
\begin{picture}(100,200)(0,0)
\put(25,0){\circle*{5}}
\multiput(0,23)(8,-8){3}{\oval(8,8)[tr]}
\multiput(8,23)(8,-8){3}{\oval(8,8)[bl]}
\multiput(0,-23)(8,8){3}{\oval(8,8)[br]}
\multiput(8,-23)(8,8){3}{\oval(8,8)[tl]}

\put(-10,26){\makebox(0,0){#1}}
\put(-10,-26){\makebox(0,0){#2}}
\end{picture}   }

\def\bs#1#2{
\begin{picture}(100,200)(0,0)
\put(0,0){\circle*{5}}
\multiput(5,0)(8,8){4}{\oval(8,8)[tl]}
\multiput(5,8)(8,8){3}{\oval(8,8)[br]}
\multiput(5,0)(8,-8){4}{\oval(8,8)[bl]}
\multiput(5,-8)(8,-8){3}{\oval(8,8)[tr]}

\put(34,30){\makebox(0,0){#1}}
\put(34,-30){\makebox(0,0){#2}}
\end{picture}   }

\def\bp#1#2{
\begin{picture}(100,200)(0,0)
\multiput(0,0)(16,0){3}{\oval(8,8)[t]}
\multiput(8,0)(16,0){3}{\oval(8,8)[b]}

\put(#2,15){\makebox(0,0){#1}}
\end{picture}   }

\def\fp#1#2{
\begin{picture}(100,200)(0,0)
\put(0,0){\line(1,0){48}}
\put(#2,15){\makebox(0,0){#1}}
\end{picture}   }

\def\be#1#2{
\begin{picture}(100,200)(0,0)
\put(25,0){\circle*{5}}
\multiput(0,23)(8,-8){3}{\oval(8,8)[tr]}
\multiput(8,23)(8,-8){3}{\oval(8,8)[bl]}
\multiput(0,-23)(8,8){3}{\oval(8,8)[br]}
\multiput(8,-23)(8,8){3}{\oval(8,8)[tl]}

\put(-13,30){\makebox(0,0){#1}}
\put(-13,-30){\makebox(0,0){#2}}
\end{picture}   }
\def\bhe#1#2{
\begin{picture}(100,200)(0,0)
\put(25,0){\circle*{5}}
\multiput(0,23)(8,-8){3}{\oval(8,8)[tr]}
\multiput(8,23)(8,-8){3}{\oval(8,8)[bl]}
\put(0,-25){\vector(1,1){27}}

\put(-13,30){\makebox(0,0){#1}}
\put(-13,-30){\makebox(0,0){#2}}
\end{picture}   }
\def\bhh#1#2{
\begin{picture}(100,200)(0,0)
\put(0,25){\vector(1,-1){27}}
\put(25,0){\circle*{5}}
\put(0,-25){\vector(1,1){27}}
\put(-13,30){\makebox(0,0){#1}}
\put(-13,-30){\makebox(0,0){#2}}
\end{picture}   }

\def\bs#1#2{
\begin{picture}(100,200)(0,0)
\put(0,0){\circle*{5}}
\multiput(5,0)(8,8){4}{\oval(8,8)[tl]}
\multiput(5,8)(8,8){3}{\oval(8,8)[br]}
\multiput(5,0)(8,-8){4}{\oval(8,8)[bl]}
\multiput(5,-8)(8,-8){3}{\oval(8,8)[tr]}

\put(34,30){\makebox(0,0){#1}}
\put(34,-30){\makebox(0,0){#2}}
\end{picture}   }

\def\bp#1#2{
\begin{picture}(100,200)(0,0)
\multiput(0,0)(16,0){3}{\oval(8,8)[t]}
\multiput(8,0)(16,0){3}{\oval(8,8)[b]}

\put(#2,15){\makebox(0,0){#1}}
\end{picture}   }

\def\fp#1#2{
\begin{picture}(100,200)(0,0)
\put(0,0){\line(1,0){48}}
\put(#2,15){\makebox(0,0){#1}}
\end{picture}   }

\def\bq{\begin{equation}}
\def\eq{\end{equation}}
\def\bqa{\begin{eqnarray}}
\def\eqa{\end{eqnarray}}
\def\bqb{\begin{eqnarray*}}
\def\eqb{\end{eqnarray*}}
\documentstyle[12pt]{article}\textwidth 16cm
\def\pr#1#2#3{ Phys. Rev. ${\bf{#1}}$ (#2) #3 }

\def\pl#1#2#3{ Phys. Lett. ${\bf{#1}}$ (#2) #3 }

\def\np#1#2#3{ Nucl. Phys. ${\bf{#1}}$ (#2) #3 }
\def\zp#1#2#3{ Z. Phys. ${\bf{#1}}$ (#2) #3 }

\def\ie{{\it i.e.\/}}
\def\eg{{\it e.g.\/}}

\def\etal{{\it et.al.\/}}

\def\L{ {\cal L }}
\def\O{ {\cal O }}

\global\nulldelimiterspace = 0pt

\def\hf{{1\over 2}}

\def\roughly#1{\mathrel{\raise.3ex
    \hbox{$#1$\kern-.75em\lower1ex\hbox{$\sim$}}}}
\def\lsim{\roughly<}
\def\gsim{\roughly>}

\def\swd{s^2_W}
\def\cwd{c^2_W}
\def\sw{s_W}
\def\cw{c_W}

\def\cot{\cos\theta}

\def\mwd{M_W^2}
\def\mw{M_W}
\def\mz{M_Z}
\def\mzd{M_Z^2}
\def\lw{\lambda_W}

\def\rd{\sqrt2}
\def\Uh{\widehat{U}}

\begin{document}
\pagenumbering{arabic}
\thispagestyle{empty}
\def\thefootnote{\fnsymbol{footnote}}
\setcounter{footnote}{1}

\begin{flushright} PM/94-28 \\ THES-TP 94/08 \\ August 1994
\end{flushright}
\vspace{2cm}
\begin{center}
{\Large\bf Unitarity Constraints for New Physics }\\
{\Large\bf Induced by dim-6 Operators}\footnote
{Work supported
by the scientific cooperation program between
 CNRS and EIE.}
 \vspace{1.5cm}  \\
{\large G.J. Gounaris$^a$, J. Layssac$^b$, J.E. Paschalis$^a$ and
F.M. Renard$^b$}
\vspace {0.5cm}  \\
$^b$Physique
Math\'{e}matique et Th\'{e}orique,
CNRS-URA 768\\
Universit\'{e} Montpellier II,
 F-34095 Montpellier Cedex 5.\\
\vspace{0.2cm}
$^a$Department of Theoretical Physics, University of Thessaloniki,\\
Gr-54006, Thessaloniki, Greece.\\

\vspace {1.5cm}

 {\bf Abstract}
\end{center}
\noindent
We compute the helicity amplitudes for boson-boson scattering
 at high energy due to the operators $\O_{B\Phi}$, $\O_{W\Phi}$ and
$\O_{UB}$,
and we derive the corresponding unitarity bounds. Thus, we
provide relations
between the couplings of these operators and the corresponding
New Physics thresholds, where either unitarity is
saturated or new degrees of freedom are excited. We compare the
results with those previously obtained for
the operators $\O_W$ and $\O_{UW}$ and we discuss their implications
for direct and indirect tests at present and future colliders.
The present treatment completes the study of the unitarity
constraints for all blind bosonic operators.

\setcounter{page}{0}

\vspace{3cm}

\def\thefootnote{\arabic{footnote}}
\setcounter{footnote}{0}
\clearpage

\section{Introduction}
At present energies where no production of any New Physics (NP)
particles has ever been observed, the search of NP effects goes
mainly through
the procedure dubbed high precision tests \cite{prtest}. It
corresponds to the hypothesis
that NP dynamics is governed by a ~characteristic scale $\Lambda_{NP}$ lying
much above the electroweak scale $v$. Therefore its observable
effects in present
high precision experiments should take the form of residual
interactions among usual
particles (leptons, quarks,
gauge bosons and possibly Higgs bosons), which are beyond those
expected in the Standard Model (SM). Such
residual interactions can be described in terms of effective
lagrangians. \par

These effective Lagrangians are constructed \cite{Buchmuller} in
terms of standard model
fields and are constrained to preserve the usual space-time and internal
symmetries of the SM. Thus $SU(2) \times U(1)$
gauge invariance is
imposed, which has the extra benefit that it tempers the
loop divergences and leads to a
decent $\Lambda_{NP}$ dependence of loop integrals involving these
 interactions \cite{DeR, Hag}. However this does not restrict
by itself the number of independent NP
operators \cite{numb}. Such a restriction is generated though from the
fact that $\Lambda_{NP} \gg v$, which hopefully means that only a few low
dimensional operators are relevant.
Since SM already includes all possible $dim=4$ terms, the NP
effects start being described by the $dim=6$ operators.\par

A restricted list of effective lagrangians has been established
on the basis of the results of the high precision tests performed
at LEP1, SLC and other low energy experiments \cite{LEP1}.
Indeed, from the absence of any
departure from the SM predictions in fermionic interactions
(at the permille level in some cases),
it seems natural to describe the NP
 effects using  operators involving only the bosonic fields
($\gamma$, Z, W, H). Imposing also CP invariance for NP, a list
of 11 independent $dim=6$  bosonic
operators has been drawn \cite{Hag}.
Four of these operators, however, affect the
 gauge boson 2-point functions at tree level and their contribution
is severely constrained by the high precision tests. Another two
depend only
on Higgs fields and do not lead to any observable effects in present
or future experiments. Consequently, we end up with five
remaining operators (the so called "blind" operators \cite{DeR}),
which are viable candidates for the ~description of observable NP
effects in the near future. These ~operators imply genuine NP gauge
boson and Higgs self interactions, involving  3-boson
{}~and multi-boson vertices.
These NP manifestations
could be observable at future machines through gauge boson pair
production as well as through  production of channels
involving Higgs bosons.\par

It has been shown that if LEP1 high precision measurements are used
to test the indirect 1-loop contributions of these operators to
the gauge boson self-energies, then the constraints obtained on
their couplings are rather mild \cite{Hag}. Therefore
considerable room exists at present, for
the observability of such interactions
at LEP2 \cite{BMTlep2} (at the level of O(0.1))
and a fortiori also at the higher energy machines
LHC \cite{GLRLHC} and NLC \cite{BMTnlc} , \cite{GRNLC},
where the sensitivity should be 10 to 100 times better.
Further restrictions on these operators may be found
by making dynamical assumptions on the origin of NP
and the additional symmetries that it might satisfy
\cite{Grosse,GRdyn, Lahanas}.\par

In this paper we discuss the validity domain of these operators
by using unitarity constraints. This is amply motivated by
the fact that at least in the well known old example of the
Fermi current-current interaction (which is also a $dim=6$
operator), unitarity considerations have
proven to be extremely ~powerful in pinpointing the correct
energy region where the underlying new physics would arise;
\ie\@ the $\mw$, $\mz$ mass domain.\par

In a similar way, the above NP local operators lead to amplitudes
involving the various gauge bosons and Higgs particles, which
approach the unitarity limit at a sufficiently high energy.
Thus, either strong
interactions will be generated at such an energy, or
new particles will be excited which will destroy the locality of
the NP operators we have started with\footnote{In fact this is
what happened to the old Fermi theory.}.
This energy value should be ~identified with the NP scale or
threshold $\Lambda_{NP}$.
So for each of the five blind operators, unitarity considerations
provide relations between their coupling constants
and the NP scale. These relations can
be used in several ways. Thus, if from some model one knows a lower
bound for the NP scale $\Lambda_{NP}$, then unitarity  can
be used to  ~obtain
upper bounds for the couplings of the various NP operators.
Or vice versa, if
an upper bound on any of these couplings is experimentally
established, then unitarity provides a
lower bound for the relevant NP threshold $\Lambda_{NP}$.
Obviously a very accurate experiment, sensitive to very small couplings,
will be able to push $\Lambda_{NP}$ to very high values.\par

In a previous paper \cite{GLRuni}
we established such relations for two of the above blind
operators. These operators were selected because they are also
invariant under custodial $SU(2)_c$ transformations.
They have the common property of generating at sufficient high
energies, strong interactions among
transverse $W_T$ states, irrespective of the Higgs mass.
This was a novel feature as compared
to the well-known case \cite{strong} of strong interactions
appearing among longitudinal
$W_L$ states in the $M_H \to \infty$ limit. We now extend this
program to the full set of blind operators. One of them ($\O_{UB}$)
can also generate strong interactions for transverse $B_T$ states,
whereas  ($\O_{W\Phi}$ and $\O_{B\Phi}$)
affect strongly the longitudinal $W_L$ and $B_L$ states also.
In this
last case though, the situation is different from the usual one
in \cite{strong}, because now strong interactions appear
even if the Higgs boson
is so light that it can possibly also be produced
\cite{GLRHiggs}.\par

We established these unitarity relations by following a 3-step procedure.
Firstly, we compute all 2-body boson-boson helicity amplitudes
involving $\gamma$, $Z$, $W$ and $H$ states, generated by any
 blind operator. Very simple expressions for these
amplitudes are obtained for c.m.\@ above $1~TeV$, by neglecting
all subleading $O(\mwd/s)$ terms \cite{GLRLHC}.
These results should, by the way, be
useful for computing the various observables in boson-boson fusion
processes at high energy colliders. Secondly, we project these high
energy amplitudes on the lowest partial waves  which give the
most stringent unitarity constraints. And thirdly, we derive the unitarity
limit for each partial wave by diagonalizing the related
matrix, thus ~getting relations between the coupling
constants and the energy scale. As explained in
\cite{GLRuni}, it is justified for our indicative purposes to
treat each blind operator separately. The results for the various
operators are discussed and compared with the indirect
constraints obtained from high precision tests, and with the sensitivities
expected at future machines. We will see that this is instructive for
scrutinizing the NP properties and identifying the sector
where they are originated.\par

The development of the paper goes through the 3 steps mentioned above.
In Sect. 2 we present the various 2-body scattering helicity amplitudes
whose high energy expressions are ~explicitly written in Appendix A and B.
In Sect. 3 we project  the partial waves and write the unitarity
constraints for the three new operators. A discussion of the results
and a comparison with other constraints is done in Sect. 4. Their
implications for the search of NP are drawn in the concluding
Sect. 5.\par

\section{Boson-boson scattering through $dim=6$ interactions}

We derive the full set of vector boson ($V=\gamma$, $Z$,
$W^{\pm}$) and Higgs
boson ($H$) scattering amplitudes in the $VV$, $HV$ and $HH$ channels,
due to the three blind operators
\bq
\O_{B\Phi} = iB^{\mu\nu}(D_\mu\Phi)^\dagger D_\nu\Phi \ \ \ \ \ ,
\ \ \ \ \ \
\eq
\bq
\O_{W\Phi} = i\overrightarrow{W}^{\mu\nu}\cdot
(D_\mu\Phi)^\dagger
\overrightarrow{\tau} D_\nu\Phi  \ \ \ \ \ \ , \ \ \ \ \ \
\eq
\bq
\O_{UB}=\frac{2}{v^2}\langle \Uh \Uh^{\dagger} -\frac{v^2}{2} \rangle
B_{\mu\nu}B^{\mu\nu}
\ \ \ \ \ \ \ \ , \ \ \ \ \
\eq
where $\overrightarrow{W}^{\mu\nu}$ is the non-abelian W field strength
and $\Uh$ is the scalar field matrix
\bq
\Uh=\bigm(\widetilde \Phi\ \ , \ \Phi\bigm) \ \ \ \ , \ \ \
\eq
\bq
\Phi=\left( \begin{array}{c}
      \phi^+ \\
{1\over\sqrt2}(v+H+i\phi^0) \end{array} \right) \ \ \ \ , \ \
\eq
$\widetilde \Phi = i\tau_2 \Phi^* $, $\langle A \rangle \equiv
Tr A$ and $v=2\mw/g_2$.\par

These processes
go through vector and Higgs boson exchange as well
as 4-~particle contact terms. The NP Lagrangian is written as
\bq
\L_{NP} = {f_B g_1\over {2M^2_W}}\O_{B\Phi} +{f_W g_2\over
{2M^2_W}}\O_{W\Phi}
+{d_B\over4}\O_{UB} +\lw \frac{g_2}{\mwd}\O_W +d\O_{UW} \ \ \ \
\ \ , \ \ \
\eq
where we have also included for later convenience the contribution
from the blind operators $\O_W$ and $\O_{UW}$
\bqa
\O_W &= & {1\over3!}\left( \overrightarrow{W}^{\ \ \nu}_\mu\times
  \overrightarrow{W}^{\ \ \lambda}_\nu \right) \cdot
  \overrightarrow{W}^{\ \ \mu}_\lambda  \ \ \
, \ \ \ \ \\[0.3cm]
\O_{UW} &= & \frac{1}{2 v^2} \, \langle \Uh \Uh^\dagger - \frac{v^2}{2}
\rangle \overrightarrow{W}_{\mu \nu}\cdot
\overrightarrow{W}^{\mu \nu}  \ \ \ \ \ , \ \ \
\eqa
analysed in
\cite{GLRuni}. The implied Feynman rules are given in Table I.\par

The explicit expressions of the helicity amplitudes
in the high energy
approximation are given in Appendix A for the $\O_{B\Phi}$ and
$\O_{W\Phi}$ contributions,
and in Appendix B for the $\O_{UB}$ case. At the tree level we are
working in, we only have
linear and quadratic terms in the new couplings $f_B$, $f_W$, $d_B$.
They grow with energy like $s$, $s^{3/2}$ or $s^2$. The leading SM
contributions can be
found in \cite{GLRLHC}, and the results for the other blind ~operators
$\O_W$ and $\O_{UW}$ are listed in \cite{GLRLHC, GLRHiggs}.
Subleading terms are
suppressed by powers of $\mwd/s$ compared to the leading ones
and they are negligible for $s\gsim 1~TeV^2$. Thus these
amplitudes are very accurate for ~energies $\gsim 1~TeV$.\par

In the $\O_{B\Phi}$ and $\O_{W\Phi}$ cases, the unitarity limits are
reached when quantities of the type $(fs/ \mwd)^2$ are large. As we are
obviously interested in scales $s\gg \mwd$, the other possible
terms, (of the form  $f^2s/ \mwd$ or
$fs/\mwd$), are negligible. This simplifies very
much the calculation, since the leading contribution arises
from just the three neutral channels  $W^-_LW^+_L$,
$Z_LZ_L$ and $HH$, which in turn means that we only
have to diagonalize a $3\times 3$  matrix. \par

In the $\O_{UB}$ case, the leading contributions can be either
of the form $d_Bs/M^2_W$ arising from $SM-\O_{UB}$ interference
in LLTT and HHTT channels, or of the form $d^2_Bs/M^2_W$ due to
purely ~transverse amplitudes involving two $\O_{UB}$ ~vertices. This fact
increases somewhat the rank of the matrix to be diagonalized in
this case, and it is similar to the situation observed in
the $\O_{UW}$ treatment \cite{GLRuni}.
We also remark that $\O_{UB}$ generates strong interactions
involving mainly $B_T$ and $H$, whereas the strong
interactions induced by $\O_{W\Phi}$
and $\O_{B\Phi}$ affect more the $W_L$, $B_L$ and $H$ states.\par

\section{Partial wave unitarity limits}

We project to partial waves the high energy helicity
amplitudes gotten in the previous
Section according to the expansion \cite{JW}
\bq
F( \lambda_1 \lambda_2 \rightarrow \lambda_3 \lambda_4)~=~16\pi
\sum_j  \left(j+\hf \right)\ D^{j*}_{\lambda_1-\lambda_2\,,
\,\lambda_3-\lambda_4}(\phi, \theta,0)
\langle \lambda_3 \lambda_4 |T^j | \lambda_1 \lambda_2 \rangle \ \ , \
\eq
for which the unitarity constraint is
\bq
|T^j| \leq 2  \ \ \ \ \ \ . \ \ \ \ \ \
\eq
The most stringent constraints come from the lowest values of the total
angular momentum j. They are obtained by separately treating the
sectors with total charge in the s-channel
$Q=2,1,0$.\par

In the $Q=2$ sector (\ie\@ the channel $W^+W^+$),  the most
{}~stringent constraint is derived from the $j=0$ amplitude
predominantly involving only $|W^+W^+LL \rangle$ state.
{}From these we obtain
\bq
|f_B| \lsim 101~{M^2_W \over{s}} \ \ \  \ , \ \ \ \ \
 |f_W| \lsim 58~ {M^2_W \over{s}} \ \  . \
\eq\par

The $Q=1$ sector contains the channels $\gamma W$, $ZW$ and $HW$
which can interact through all three types of operators. In the
case of $\O_{W\Phi}$ and $\O_{B\Phi}$, the ~most important
$j=0$ partial amplitude relates the
6 states ($|ZW\pm\pm \rangle$, $|ZWLL \rangle $, $|\gamma
W\pm\pm \rangle $, $|HWL \rangle $) giving
the unitarity bounds are
\bq |f_B| \lsim 98~{M^2_W \over{s}} \ \ \ \  , \ \ \ \
|f_W| \lsim 54~{M^2_W\over{s}} \ \ . \
\eq
No $j=0$ amplitude  involving $\O_{UB}$ appears in the $Q=1$
sector. So, we have to
consider the $j=1$ amplitudes. However these only contain terms
linear in $d_B$, so that the bound is rather weak
\bq |d_B| \lsim 768~{M^2_W\over{s}} \ \ \ \ \ \ \ .\ \ \ \
\eq\par

The $Q=0$ (neutral) sector is the richest one. $\O_{B\Phi}$ and
$\O_{W\Phi}$
contribute to the 12 $j=0$ states, ($|\gamma\gamma\pm\pm
{}~\rangle $, $ |\gamma Z\pm\pm \rangle $,
$|HZ L \rangle $, $|W^-W^+\pm\pm \rangle $, $|W^-W^+LL \rangle $,
$|ZZ\pm\pm\rangle $, $|ZZLL \rangle $,
$|HH\rangle $). From the diagonalization of this $j=0$ amplitude
we get
\bq
|f_B| \lsim 98~{M^2_W\over{s}}  \ \ \ \ \ , \ \ \ \
\  |f_W| \lsim 31~{M^2_W \over{s}} \ \ \ \ . \
\eq
For the
$\O_{UB}$ case, the 9 states ($|\gamma\gamma\pm\pm \rangle$,
$|\gamma Z\pm\pm\rangle $,
$|ZZ\pm\pm \rangle $, $|ZZ LL \rangle $, $|W^-W^+LL \rangle $,
$|HH \rangle $) are
the ones which predominantly participate in the $j=0$ ~partial amplitude.
 From its diagonalization we get
\bq
{\alpha d_B s \over {4M^2_W}}(\sqrt{33d^2_B+16d_B+8}-d_B)
\lsim 1 \ \ \ \ \ \ \ \ , \ \ \
\eq
whose numerical solution for  $s \gsim 1~ TeV^2$ is
\bq
-236~{M^2_W\over{s}}~+~1070~{M^3_W\over{s^{3/2}}}~ \lsim ~d_B ~\lsim
{}~ 192~{M^2_W\over{s}}~-~
1123~{M^3_W\over{s^{3/2}}}  \ \ \ \ .\ \ \
\eq
For $\alpha s/\mwd \gg 1~TeV^2$ (i.e.  $ s\gsim 10 ~TeV^2$) this result
simplifies to
\bq |d_B| \lsim 180~{\mwd \over s}\ \ \ \ \ \ \ . \ \ \ \
\eq.

\section{Panorama of unitarity constraints.}

The most stringent results found above are
\bq
|f_B| \leq 98{M^2_W\over{s}} \ \ \ \ \ , \ \ \ \ \
 \ |f_W| \leq 31{M^2_W \over{s}} \ \ \ \ ,\
 \eq
and eqs.\@ (16, 17) for $d_B$. Together with these we recall
  the corresponding unitarity
constraints on $\O_W$ and $\O_{UW}$ derived in \cite{GLRuni}
\bq
|\lambda_W| \lsim 19~{M^2_W \over s} \ \ \ \ \ , \ \ \ \ \ \ \
\ |d| \lsim 17.6~{M^2_W\over{s}}+2.43
{M_W\over{\sqrt{s}}} \ \ \ \ \ \ \ \ \ ; \ \ \
\eq
(compare (6)). We now discuss the unitarity bounds obtained
for all five blind operators.\par

The bound on $f_B$ is somewhat weaker than the other ones,
because of its normalization through the smaller value of
$g_1$ rather than that of $g_2$. Recall the definition of these
couplings given in (6).
The bound (16, 17) for $d_B$ is also somewhat weaker than the one for
d in (19). This can be understood from the definitions (3, 8) of
the ~corresponding operators, by remarking
that there is no $HWW$ coupling through $d_B$,
while  the $HZZ$ coupling induced by $d_B$ is weaker than the
$d$ induced one by a factor of $s^2_W/c^2_W$; (note that the role
of ZZ and $\gamma\gamma$ are interchanged when passing from d to
$d_B$). We also remark that this $d_B$ versus $d$ comparison would
have been more
striking if we had not used the factors of two in the ~definitions
(3, 8). \par

In practice, assuming a certain value for the NP scale $s=\Lambda^2_{NP}$,
one deduces upper bounds for the various couplings. For example
if $\Lambda_{NP} = 1 ~TeV$ one obtains
\bq
|f_B| \lsim 0.6 \ \ \ , \ \ \ |f_W| \leq 0.2 \ \ \ , \ \ \
 -0.8 \lsim d_B \lsim 0.6 \ \ , \ \
\eq
\bq |\lambda_W| \lsim 0.12 \ \ \ \ , \ \ \  \ |d| \lsim 0.3
\ \ \ \ . \ \ \eq
These relations provide a feeling of how sensitive the various
couplings are to unitarity constraints.
Conversely, from the expected sensitivities to these couplings
at future colliders, one can deduce the achievable
lower bounds for the NP scale $\Lambda_{NP}$ at these machines;
\ie\@ the lower bound for either the generation of new
strong interactions, or the production of new particles. For example
at NLC (0.5 TeV) where the observability limits
can be written as  \cite{BMTnlc}
\bq
|f_B| \gsim 0.012 \ \ \ \ \ , \ \ \ \ \ \ \ \ |f_W| \gsim 0.006
\ \ \ \ , \  \ \eq
we expect to be sensitive to NP scales satisfying
\bq
\Lambda_{NP}(f_B) \lsim 7~ TeV \ \ \ \ \ , \ \ \ \ \ \
\ \ \Lambda_{NP}(f_W) \lsim 6~ TeV\ \ \ \ \  ,\  \ \
\eq
to be compared to
\bq
\Lambda_{NP}(\lambda_W) \lsim 4~ TeV\ \ \ \ \ \ \ , \ \ \ \
\eq
obtained from $|\lambda_W| \gsim 0.008$ \cite{BMTnlc}.\par

We next turn to the Higgs sector.  For $\O_{UW}$, the highest
sensitivity  $|d| \gsim 0.001$ was obtained  from
$\gamma\gamma \to H$ production
in laser backscattering experiments \cite{GLRHiggs}.  This implies
\bq \Lambda_{NP}(d) \leq 30~TeV\ \ \ \ \ .\ \ \ \  \eq
In the $\O_{UB}$ case the $\gamma\gamma \to H$ production rate
is enhanced by the factor $c^2_W/s^2_W$. From statistics one
then expects an increase in  sensitivity
by a factor 3, which means $|d_B| \gsim 3.10^{-4}$, and from
eq(16, 17)
\bq \Lambda_{NP}(d_B) \lsim 60~ TeV \ \ \ \ .\ \ \  \eq\par

\section{Implications for New Physics searches}

We have established unitarity constraints for effective interactions
which turn out to have many implications. They give
relations between the coupling constants of each blind operator
and the related NP scale at
which new phenomena should appear. For example, assuming that the NP
scale (or a lower bound of this) is known, one obtains an
upper bound for the various
couplings. Thus if \eg\@ $\Lambda_{NP} \gsim 1~TeV^2$, then
$|f_B| \lsim 0.6$, $|f_W|\lsim 0.2$,
$|\lambda_W|\lsim 0.12$,
$-0.8\lsim d_B\lsim 0.6$ and $|d|\lsim 0.3$.
Such bounds are quite interesting.
They lie in the same range as those obtained
by calculating the indirect 1-loop effects of the blind operators
using the LEP1 constraints \cite{DeR, Hag}.
However, when doing such  1-loop computations with blind
operators, one
should remember that the NP contributions in the energy range
$s \gsim \Lambda^2_{NP}$ are actually ignored, while the lower
energies contribute.
For the validity of such calculations, it should therefore be  checked, a
posteriori, whether strong ~interactions have not already been
developed in the energy range affecting the result. It
is obvious that in the later case the perturbative treatment
would be questioned. Moreover, the only way to justify ignoring
the contributions from a strongly interacting energy regime is
to assume that somehow the theory softens there. More concretely,
one should worry whether such a treatment is justified in case
the values of the coupling constants obtained and the NP scales
{}~assumed, violate our unitarity relations.\par

Another aspect of our unitarity constraints is to associate in a
{}~simple way the NP scale to the observability limits which could
be established
for each effective interaction at future colliders. In that way one
can clearly see that LEP2 experiments could ~explore the TeV
range $\Lambda_{NP}$, while the LHC and NLC ones should be
sensitive to NP at
scales up to several tens of TeV.\par

It is then interesting to examine more carefully the structure of the
effective operators and the nature of the NP effects involved.
In the former cases \cite{GLRuni} of $\O_W$ and $\O_{UW}$,
as well as in the $\O_{UB}$
case treated in this paper, strong interactions appear among transverse
gauge boson  ($W_T$ , $Z_T$ , $\gamma_T$) and Higgs
states. The two other
operators $\O_{W\Phi}$ and $\O_{B\Phi}$ generate strong interactions
mainly among longitudinal $W_L$, $B_L$ and Higgs states. Note that contrary
to the SM case for which strong $W_L$ interactions appear in the $M_H \to
\infty$ limit, here it is not necessary for the Higgs mass to be large.
These strong interactions appear even when the Higgs boson is light,
and this light Higgs is itself strongly coupled to either $W_T$, $B_T$ or
$W_L$, $B_L$ states. This means that each class of effective operators
has a different implication about the NP properties and their origin.
It is then extremely useful to disentangle these various possible NP
manifestations in experimental measurements, or to precisely
determine the observability limits for each of these new interactions
separately. This will allow to test the NP pictures that one can have
in mind, or at least to discriminate among the various sectors
from which NP can originate.\par
\null\par

\vspace{2cm}

{\Large \bf Acknowledgements}\vspace{0.5cm}\par
 One of us (F.M.R.) wishes to thank the Department of Theoretical
 Physics of the
 University
of Thessaloniki for the warm hospitality
as well as the kind help that he
 received
 during his
stay and for the creative atmosphere that was generated during
the preparation of this paper.
\newpage

{\large \bf Table I: Feynman rules for $\O_{W\Phi}$,
$\O_{B\Phi}$ and $\O_{UB}$
interactions}\par
\vspace{1cm}

{}From the NP Lagrangian of eq(6) in the unitary-gauge, one derives
the following expressions for the 3- and 4- body vertices
\def\cw{\cos\theta_W}
\def\sw{\sin\theta_W}
\def\cwd{\cos^2\theta_W}
\def\swd{\sin^2\theta_W}
\def\fw{g_2\ f_W}
\def\fb{g_1\ f_B}
\def\mw{M_W}
\def\mwd{M_W^2}

\baselineskip=18pt
\par
\begin{picture}(100,100)(0,0)
\put(10,50) {\bp{$A_\mu$}{24}}
\put(20,40){\makebox(0,0){$p$}}
\put(25,40){\vector(1,0){15}}
\put(75,75){\vector(-1,-1){15}}
\put(75,25){\vector(-1,1){15}}
\put(56,50) {\bs{$\ \ \ \ W^+_\alpha$}{$\ \ \ \ W^-_\beta$}}
\end{picture}
\vspace{-65pt}\null\par
\hangindent=7cm
\hspace*{6.5cm}$\displaystyle {i\over2}(\fw\sw +\fb\cw)
(p_\beta g_{\mu\alpha}-p_\alpha g_{\mu\beta})$
\vspace{2cm}
\par
\begin{picture}(100,100)(0,0)
\put(10,50) {\bp{$Z_\mu$}{24}}
\put(20,40){\makebox(0,0){$p$}}
\put(25,40){\vector(1,0){15}}
\put(56,50) {\bs{$\ \ \ \ W^+_\alpha$}{$\ \ \ \ W^-_\beta$}}
\put(75,75){\vector(-1,-1){15}}
\put(70,80){\makebox(0,0){$p_1$}}
\put(75,25){\vector(-1,1){15}}
\put(70,20){\makebox(0,0){$p_2$}}
\end{picture}
\vspace{-75pt}\null\par
\hangindent=7cm
\hspace*{6.5cm}$\displaystyle {-i\fw\over2\cw}
 \{p_{1\beta} g_{\alpha\mu}-p_{1\mu} g_{\alpha\beta}
  -p_{2\alpha} g_{\beta\mu}+p_{2\mu} g_{\alpha\beta}\}$\\
$\displaystyle +{i\over2}(\fw\cw -\fb\sw)
(p_\beta g_{\mu\alpha}-p_\alpha g_{\mu\beta})$
\vspace{2cm}
\par
\begin{picture}(100,100)(0,0)
\put(10,50) {\fp{$H_0$}{24}}
\put(20,40){\makebox(0,0){$p$}}
\put(25,40){\vector(1,0){15}}
\put(56,50) {\bs{$\ \ \ \ W^+_\alpha$}{$\ \ \ \ W^-_\beta$}}
\put(75,75){\vector(-1,-1){15}}
\put(70,80){\makebox(0,0){$p_1$}}
\put(75,25){\vector(-1,1){15}}
\put(70,20){\makebox(0,0){$p_2$}}
\end{picture}
\vspace{-65pt}\null\par
\hangindent=7cm
\hspace*{6.5cm}$\displaystyle {i\fw\over2M_W}
 \{ g_{\alpha\beta}\ p\ .\ (p_1+p_2)-p_\alpha\ p_{1\beta}
     -p_\beta\ p_{2\alpha}\} $
\vspace{2cm}
\par
\begin{picture}(100,100)(0,0)
\put(10,50) {\fp{$H_0$}{24}}
\put(20,40){\makebox(0,0){$p$}}
\put(25,40){\vector(1,0){15}}
\put(56,50) {\bs{$\ \ \ \ Z_\alpha$}{$\ \ \ \ Z_\beta$}}
\put(75,75){\vector(-1,-1){15}}
\put(70,80){\makebox(0,0){$p_1$}}
\put(75,25){\vector(-1,1){15}}
\put(70,20){\makebox(0,0){$p_2$}}
\end{picture}
\vspace{-75pt}\null\par
\hangindent=7cm
\hspace*{6.5cm}$\displaystyle {i\over2M_W}\bigm[(\fw+\fb \tan \theta_W)
 \{ g_{\alpha\beta}\ p\ .\ (p_1+p_2)-p_\alpha\ p_{1\beta}
     -p_\beta\ p_{2\alpha}\}$\\
$\displaystyle -\ 4\ d_B\ g_2\ \swd
 \{ g_{\alpha\beta}\ p_1\ .\ p_2 - p_{1\beta}\ p_{2\alpha}\}\bigm] $
\vspace{2cm}
\par
\begin{picture}(100,100)(0,0)
\put(10,50) {\fp{$H_0$}{24}}
\put(20,40){\makebox(0,0){$p$}}
\put(25,40){\vector(1,0){15}}
\put(56,50) {\bs{$\ \ \ \ A_\mu$}{$\ \ \ \ Z_\alpha$}}
\put(75,75){\vector(-1,-1){15}}
\put(70,80){\makebox(0,0){$p_1$}}
\put(75,25){\vector(-1,1){15}}
\put(70,20){\makebox(0,0){$p_2$}}
\end{picture}
\vspace{-75pt}\null\par
\hangindent=7cm
\hspace*{6.5cm}$\displaystyle {i\over2M_W}\bigm[(\fw \tan\theta_W-\fb)
 \{ g_{\mu\alpha}\ (p\ .\ p_1)-p_\mu\ p_{1\alpha} \}$\\
$\displaystyle +\ 4\ d_B\ g_2\ \sw\cw
 \{ g_{\mu\alpha}\ p_1\ .\ p_2 - p_{2\mu}\ p_{1\alpha}\}\bigm] $
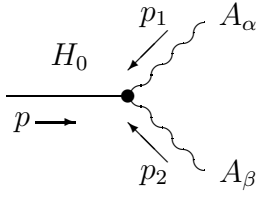
\vspace{2cm}
\par
\begin{picture}(100,100)(0,0)
\put(10,50) {\fp{$H_0$}{24}}
\put(20,40){\makebox(0,0){$p$}}
\put(25,40){\vector(1,0){15}}
\put(56,50) {\bs{$\ \ \ \ A_\alpha$}{$\ \ \ \ A_\beta$}}
\put(75,75){\vector(-1,-1){15}}
\put(70,80){\makebox(0,0){$p_1$}}
\put(75,25){\vector(-1,1){15}}
\put(70,20){\makebox(0,0){$p_2$}}
\end{picture}
\vspace{-75pt}\null\par
\hangindent=7cm
\hspace*{6.5cm}$\displaystyle -\ {2i\ d_B\ g_2\ \cwd\over\mw}
 \{ g_{\alpha\beta}\ p_1\ .\ p_2 - p_{1\beta}\ p_{2\alpha}\} $
%
\def\cw{\cos\theta_W}
\def\sw{\sin\theta_W}
\def\fw{g_2\ f_W}
\def\fb{g_1\ f_B}

\vspace{2cm}
\par
\begin{picture}(100,100)(0,0)
\put(30,50) {\be{$ W^+_\alpha$}{$ W^+_\gamma$}}
\put(31,40){\makebox(0,0){$p_3$}}
\put(33,30){\vector(1,1){15}}
\put(31,60){\makebox(0,0){$p_1$}}
\put(33,70){\vector(1,-1){15}}
\put(56,50) {\bs{$\ \ \ \ W^-_\beta$}{$\ \ \ \ W^-_\delta$}}
\put(90,70){\vector(-1,-1){15}}
\put(95,60){\makebox(0,0){$p_2$}}
\put(90,30){\vector(-1,1){15}}
\put(95,40){\makebox(0,0){$p_4$}}
\end{picture}
\vspace{-65pt}\null\par
\hangindent=5.5cm
\hspace*{5.cm}$\displaystyle i\fw\ g_2
 \{2 g_{\alpha\gamma} g_{\beta\delta}-g_{\alpha\beta} g_{\gamma\delta}
   -g_{\alpha\delta} g_{\beta\gamma}\}$
\vspace{2cm}
\par
\begin{picture}(100,100)(0,0)
\put(30,50) {\be{$ W^+_\alpha$}{$ A_\gamma$}}
\put(31,40){\makebox(0,0){$p_3$}}
\put(33,30){\vector(1,1){15}}
\put(31,60){\makebox(0,0){$p_1$}}
\put(33,70){\vector(1,-1){15}}
\put(56,50) {\bs{$\ \ \ \ W^-_\beta$}{$\ \ \ \ Z_\delta$}}
\put(90,70){\vector(-1,-1){15}}
\put(95,60){\makebox(0,0){$p_2$}}
\put(90,30){\vector(-1,1){15}}
\put(95,40){\makebox(0,0){$p_4$}}
\end{picture}
\vspace{-65pt}\null\par
\hangindent=5.5cm
\hspace*{5.cm}$\displaystyle {-i\fw\over2}\sw(g_2\cw+g_1\sw)
 \{2 g_{\alpha\beta} g_{\gamma\delta}-g_{\alpha\gamma} g_{\beta\delta}
   -g_{\alpha\delta} g_{\beta\gamma}\}$
\newpage
\vspace{2cm}
\par
\begin{picture}(100,100)(0,0)
\put(30,50) {\be{$ W^+_\alpha$}{$ Z_\gamma$}}
\put(31,40){\makebox(0,0){$p_3$}}
\put(33,30){\vector(1,1){15}}
\put(31,60){\makebox(0,0){$p_1$}}
\put(33,70){\vector(1,-1){15}}
\put(56,50) {\bs{$\ \ \ \ W^-_\beta$}{$\ \ \ \ Z_\delta$}}
\put(90,70){\vector(-1,-1){15}}
\put(95,60){\makebox(0,0){$p_2$}}
\put(90,30){\vector(-1,1){15}}
\put(95,40){\makebox(0,0){$p_4$}}
\end{picture}
\vspace{-75pt}\null\par
\hangindent=5.5cm
\hspace*{5.cm}$\displaystyle -i\fw\cw(g_2\cw+g_1\sw)
 \{2 g_{\alpha\beta} g_{\gamma\delta}-g_{\alpha\delta} g_{\gamma\beta}
   -g_{\alpha\gamma} g_{\beta\delta}\}$
\vspace{2cm}
\par
\begin{picture}(100,100)(0,0)
\put(30,50) {\bhe{$ W^+_\alpha$}{$H$}}
\put(31,40){\makebox(0,0){$p_3$}}
\put(33,30){\vector(1,1){15}}
\put(31,60){\makebox(0,0){$p_1$}}
\put(33,70){\vector(1,-1){15}}
\put(56,50) {\bs{$\ \ \ \ W^-_\beta$}{$\ \ \ \ A_\gamma$}}
\put(90,70){\vector(-1,-1){15}}
\put(95,60){\makebox(0,0){$p_2$}}
\put(90,30){\vector(-1,1){15}}
\put(95,40){\makebox(0,0){$p_4$}}
\end{picture}
\vspace{-85pt}\null\par
\hangindent=5.5cm
\hspace*{5.cm}$\displaystyle {ig^2_2 f_W\over2M_W}\sw
 (p_{3\alpha} g_{\beta\gamma}-p_{3\beta} g_{\alpha\gamma})$\\
$\displaystyle +\ {ig_2\over2M_W}(\fw\sw+\fb\cw)
(p_{4\beta} g_{\alpha\gamma}-p_{4\alpha} g_{\beta\gamma})$
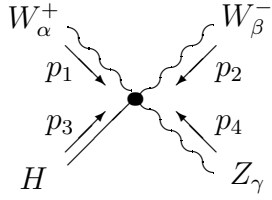
\vspace{2cm}
\par
\begin{picture}(100,100)(0,0)
\put(30,50) {\bhe{$ W^+_\alpha$}{$H$}}
\put(31,40){\makebox(0,0){$p_3$}}
\put(33,30){\vector(1,1){15}}
\put(31,60){\makebox(0,0){$p_1$}}
\put(33,70){\vector(1,-1){15}}
\put(56,50) {\bs{$\ \ \ \ W^-_\beta$}{$\ \ \ \ Z_\gamma$}}
\put(90,70){\vector(-1,-1){15}}
\put(95,60){\makebox(0,0){$p_2$}}
\put(90,30){\vector(-1,1){15}}
\put(95,40){\makebox(0,0){$p_4$}}
\end{picture}
\vspace{-95pt}\null\par
\hangindent=5.5cm
\hspace*{5.cm}$\displaystyle {i\fw\over2M_W}\bigm\{ (g_2\cw+g_1\sw)$\\
$\displaystyle
 [(p_{2\alpha} g_{\beta\gamma}-p_{2\gamma} g_{\alpha\beta}
  -p_{1\beta} g_{\alpha\gamma}+p_{1\gamma} g_{\alpha\beta}]
+g_1\sw\ [p_{3\beta} g_{\alpha\gamma}-p_{3\alpha} g_{\beta\gamma}]\bigm\}
 +{ig_1\over2M_W}\ (\fw\cw-\fb\sw)
[p_{4\beta} g_{\alpha\gamma}-p_{4\alpha} g_{\beta\gamma}] $
%
%
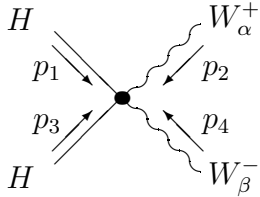
\vspace{2cm}
\par
\begin{picture}(100,100)(0,0)
\put(30,50) {\bhh{$H$}{$H$}}
\put(31,40){\makebox(0,0){$p_3$}}
\put(33,30){\vector(1,1){15}}
\put(31,60){\makebox(0,0){$p_1$}}
\put(33,70){\vector(1,-1){15}}
\put(56,50) {\bs{$\ \ \ \ W^+_\alpha$}{$\ \ \ \ W^-_\beta$}}
\put(90,70){\vector(-1,-1){15}}
\put(95,60){\makebox(0,0){$p_2$}}
\put(90,30){\vector(-1,1){15}}
\put(95,40){\makebox(0,0){$p_4$}}
\end{picture}
\vspace{-75pt}\null\par
\hangindent=5.5cm
\hspace*{5.cm}$\displaystyle {-i\fw\over4M_W^2}\ g_1 \ \bigm\{
 p_{2\beta}\ p_{1\alpha}-p_1\ .\ p_2\ g_{\alpha\beta}
  +p_{1\beta}\ p_{4\alpha}-p_1\ .\ p_4\ g_{\alpha\beta}
  +p_{2\beta}\ p_{3\alpha}-p_2\ .\ p_3\ g_{\alpha\beta}
  +p_{3\beta}\ p_{4\alpha}-p_4\ .\ p_3\ g_{\alpha\beta}\bigm\} $
\newpage
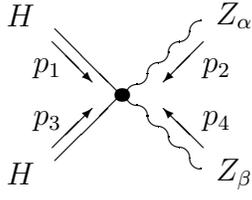
\vspace{2cm}
\par
\begin{picture}(100,100)(0,0)
\put(30,50) {\bhh{$H$}{$H$}}
\put(31,40){\makebox(0,0){$p_3$}}
\put(33,30){\vector(1,1){15}}
\put(31,60){\makebox(0,0){$p_1$}}
\put(33,70){\vector(1,-1){15}}
\put(56,50) {\bs{$\ \ \ \ Z_\alpha$}{$\ \ \ \ Z_\beta$}}
\put(90,70){\vector(-1,-1){15}}
\put(95,60){\makebox(0,0){$p_2$}}
\put(90,30){\vector(-1,1){15}}
\put(95,40){\makebox(0,0){$p_4$}}
\end{picture}
\vspace{-95pt}\null\par
\hangindent=5.5cm
\hspace*{5.cm}$\displaystyle{-i\over4M_W^2}\bigm[(\fw\cw+\fb\sw)
(g_1\cw+g_2\sw)$\\
$\displaystyle  \bigm\{
 p_{2\beta}\ p_{3\alpha}-p_3\ .\ p_2\ g_{\alpha\beta}
  +p_{3\beta}\ p_{4\alpha}-p_3\ .\ p_4\ g_{\alpha\beta}
  +p_{2\beta}\ p_{1\alpha}-p_2\ .\ p_1\ g_{\alpha\beta}
  +p_{1\beta}\ p_{4\alpha}-p_1\ .\ p_4\ g_{\alpha\beta}\bigm\}$\\
$\displaystyle +\ 16\ d_B\ g_2^2\ \swd
 \{ g_{\alpha\beta}\ p_2\ .\ p_4 - p_{2\beta}\ p_{4\alpha}\}\bigm] $
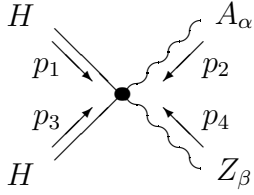
\vspace{2cm}
\par
\begin{picture}(100,100)(0,0)
\put(30,50) {\bhh{$H$}{$H$}}
\put(31,40){\makebox(0,0){$p_3$}}
\put(33,30){\vector(1,1){15}}
\put(31,60){\makebox(0,0){$p_1$}}
\put(33,70){\vector(1,-1){15}}
\put(56,50) {\bs{$\ \ \ \ A_\alpha$}{$\ \ \ \ Z_\beta$}}
\put(90,70){\vector(-1,-1){15}}
\put(95,60){\makebox(0,0){$p_2$}}
\put(90,30){\vector(-1,1){15}}
\put(95,40){\makebox(0,0){$p_4$}}
\end{picture}
\vspace{-85pt}\null\par
\hangindent=5.5cm
\hspace*{5.cm}$\displaystyle{-i\over4M_W^2}\bigm[
(\fw\sw-\fb\cw)(g_1\cw+g_2\sw)$\\
$\displaystyle  \bigm\{
 p_{2\beta}\ p_{3\alpha}-p_3\ .\ p_2\ g_{\alpha\beta}
  +p_{2\beta}\ p_{1\alpha}-p_2\ .\ p_1\ g_{\alpha\beta}\bigm\}\\$
$\displaystyle -\ 16\ d_B\ g_2^2\ \cw\sw
 \{ g_{\alpha\beta}\ p_2\ .\ p_4 - p_{2\beta}\ p_{4\alpha}\}\bigm] $
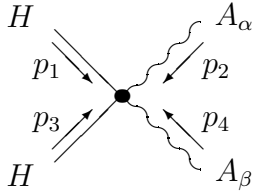
\vspace{2cm}
\par
\begin{picture}(100,100)(0,0)
\put(30,50) {\bhh{$H$}{$H$}}
\put(31,40){\makebox(0,0){$p_3$}}
\put(33,30){\vector(1,1){15}}
\put(31,60){\makebox(0,0){$p_1$}}
\put(33,70){\vector(1,-1){15}}
\put(56,50) {\bs{$\ \ \ \ A_\alpha$}{$\ \ \ \ A_\beta$}}
\put(90,70){\vector(-1,-1){15}}
\put(95,60){\makebox(0,0){$p_2$}}
\put(90,30){\vector(-1,1){15}}
\put(95,40){\makebox(0,0){$p_4$}}
\end{picture}
\vspace{-55pt}\null\par
\hangindent=5.5cm
\hspace*{5.cm}
$\displaystyle
 -\ {4i\ d_B\ g_2^2\ \cwd\over\mwd}
 \{ g_{\alpha\beta}\ p_2\ .\ p_4 - p_{2\beta}\ p_{4\alpha}\} $

\newpage
\def\ep#1#2{(\epsilon_{#1}\epsilon_{#2})}
\def\dh#1{ {1\over D_H({#1})} }
\def\nh#1{  D_H({#1}) }
\def\co{\biggm[}
\def\cf{\biggm]}

\def\mw{M_W}
\def\mwd{M_W^2}
\def\mz{M_Z}
\def\mzd{M_Z^2}
\def\ct{\cos\theta}
\def\ctp{(\cos\theta+1)\ }
\def\ctm{(\cos\theta-1)\ }
\def\cw{c_W}
\def\sw{s_W}
\def\cwd{c_W^2}
\def\swd{s_W^2}
\def\cs{c_W s_W}
\def\rs{\sqrt s}
\def\rd{\sqrt2}
\def\g{\gamma}
\def\fbd{g^2_1\ f^2_B}
\def\fwd{g^2_2\ f^2_W}
\def\feb{e\ g_1 f_B}
\def\few{e\ g_2 f_W}
\def\fbw{g_1 f_B\ g_2 f_W}

\renewcommand{\theequation}{A.\arabic{equation}}
\setcounter{equation}{0}
\setcounter{section}{0}
{\large \bf Appendix A : Helicity amplitudes for Boson-Boson fusion
processes at High Energy due to $\O_{W\Phi}$ and
$\O_{B\Phi}$ interactions}\par
\vspace{1cm}
Below we give the amplitudes for the processes that do not vanish
at high energies.
\renewcommand{\thesubsection}{A.\arabic{subsection}}
\subsection{4- gauge boson processes}
In general there are 81  helicity amplitudes $F_{\lambda \lambda'
 \mu \mu'}$
for each vector boson-vector boson fusion
process  $V_1(\lambda) V_2(\lambda') \to
V_3(\mu) V_4(\mu')$. Taking into account parity conservation,
(which is valid at tree
level for the self-boson interactions contained in SM and the
operators considered) we obtain
\bq
F_{\lambda\lambda'\mu\mu'}(\theta)~=~F_{-\lambda-\lambda'-\mu-\mu'}
(\theta)~(-1)^{\lambda - \lambda'- \mu + \mu'}
\ \ \  , \ \ \ \ \eq
\noindent
which reduces the number of independent amplitudes to 41. In
specific channels this  number is
further reduced due to \eg\@ to the absence of helicity zero states for
 photons,  the symmetrization
for identical particles, charge conjugation relations, etc. Here and
below $\theta$ is the c.m.\@ angle between $V_1$ and $V_3$. The
normalization of the amplitudes is defined by noting that the
differential cross section in c.m.\@ is given by
\bq
{d\sigma(\lambda \lambda' \mu \mu')\over dcos\theta} = C
 |F_{\lambda \lambda'  \mu
 \mu'}|^2 \ \ \ \ , \ \ \ \
\eq
 where the coefficient
\bq
C = {1\over32\pi s}\,{p_{34}\over p_{12}}
\eq
includes \underline{no spin average}.  This later choice is
motivated by the fact that, inside the proton, different vector
boson distribution functions  occur for different initial
 helicity
states. Finally  $p_{12}$, $p_{34}$ in (B.3)
denote the c.m.\@ momenta of the initial and final boson pairs
respectively. \par

As in \cite{GLRLHC}, for $s \gsim ~1TeV^2$ simple and very
accurate expressions for the
$\O_{W\Phi}$ and $\O_{B\Phi}$ contributions to the boson
amplitudes are obtained by neglecting terms of order
$O(\mwd/s)$ with respect to the leading ones. The independent
amplitudes for the various ~processes are given below as coefficients
of the specified products of coupling constants\footnote{The terms
linear in the coupling constants $f_B$ and $f_W$ coming from the
diagrams which do not involve the Higgs boson have also been computed
by I.Kuss \cite{Kuss}. }. The charge
assignment of $W$ is omitted ~whenever it is irrelevant.
\vspace{1.5cm}\par
\newpage
%
%
%
\centerline{ \fbox{$\g Z\to W^-W^+$}  }
$\fbd$
\bqa
F_{+--+}&=&\cw\ F_{+0-0}={\ctm\cs s\over32\mwd}  \nonumber\\[0.2cm]
  F_{+-00}&=&F_{++--}=-{1\over2}F_{++00}=
       -{\cs s\over16\mwd}\nonumber\\[0.2cm]
     F_{+-+-}&=&-\ \cw\ F_{+00-}=-{\ctp\cs s\over32\mwd}
\eqa\\

$\fwd$
\bqa
F_{+--+}&=&-{\swd\over \cw}
 F_{+0-0}={\ctm s^3_W s\over32\cw\mwd}\nonumber\\
     F_{+-00}&=&-{1\over2}F_{++00}={\swd\over2} F_{++++}=
             -{s^3_W s\over16\cw\mwd} \nonumber\\[0.2cm]
     F_{+-+-}&=&{\swd\over \cw}
 F_{+00-}=-\ {\ctp s^3_W s\over32\cw\mwd} \nonumber\\[0.2cm]
     F_{++--}&=&{(\cwd-3)\ \sw s\over16\cw\mwd }
\eqa\\

$\feb$
\bqa  F_{+0-0}&=&-\ {\ctm\cw s\over16\mwd\sw} \nonumber\\[0.2cm]
      F_{+00-}&=&-{\ctp\cw s\over16\mwd\sw}\ \ \ \ \ \
      F_{++00}={(3-4\cwd)\ s\over8\mwd\sw}
\eqa\\

$\few$
\bqa
F_{+0-0}&=&-{\ctm s\over16\mwd} \ \ \ \ \ \
      F_{+00-}=-{\ctp s\over16\mwd}\nonumber\\[0.2cm]
      F_{++00}&=&{(1-4\cwd)\ s\over8\cw\mwd}
\eqa\\

$ \fbw$
\bqa
F_{+--+}&=&{\ctm\swd s\over16\mwd} \ \ \ \ \ \
      F_{+-00}=-\swd F_{++++}=-{\swd s\over8\mwd}\nonumber\\[0.2cm]
      F_{+-+-}&=&-{\ctp\swd s\over16\mwd} \ \ \ \ \ \
      F_{+0-0}={\ctm\ (1-2\cwd)\  s\over32\cw\mwd}\nonumber\\[0.2cm]
      F_{+00-}&=&{\ctp\ (1-2\cwd)\ s\over32\cw\mwd} \ \ \ \ \ \
      F_{++--}= -{1\over2}F_{++00}={\cwd s\over8\mwd}
\eqa\\

\vspace{1.5cm}
\centerline{ \fbox{$\g W\to ZW$}  }
$\fbd$
\bqa
    F_{+0+0}&=&-F_{++--}=-{1\over2} F_{+0-0}=
           {\ctm\cs s\over32\mwd}  \nonumber\\[0.2cm]
     F_{+00-}&=&-{1\over\cw}F_{+-+-}={\ctp\sw s\over32\mwd}
\nonumber\\[0.2cm]
     F_{++++}&=&-\cw F_{++00}={\cs s\over16\mwd}
\eqa\\

$\fwd$
\bqa
 F_{+-+-}&=&{\swd\over\cw}
 F_{+00-}=-{\ctp s^3_W s\over32\cw\mwd}\nonumber\\[0.2cm]
     F_{+--+}&=&{1\over\swd}F_{+0-0}=-{2\over\swd}F_{+0+0}=
             -{\ctm\sw s\over16\cw\mwd} \nonumber\\[0.2cm]
     F_{++00}&=&{\cwd\over\swd} F_{++++}={\cs s\over16\mwd} \nonumber\\
     F_{++--}&=&{(\cwd-3)\ctm \sw s\over 32\cw\mwd }
\eqa\\

$\feb$
\bqa
F_{++00}&=&{\cw s\over8\mwd\sw} \ \ \ \ \
      F_{+00-}=-{\ctp\cw s\over16\mwd\sw} \nonumber\\[0.2cm]
      F_{+0-0}&=-&{\ctm (3-4\cwd)\ s\over16\mwd\sw} \ \ \ \
\eqa\\

$\few$
\bqa
F_{++00}&=&{ s\over8\mwd} \ \ \ \ \ \ \
      F_{+00-}=-{\ctp s\over16\mwd} \nonumber\\[0.2cm]
      F_{+0-0}&=&-{\ctm(1-4\cwd)\ s\over16\mwd\cw} \ \ \ \ \
\eqa\\

$ \fbw$
\bqa
F_{+--+}&=&{1\over2\cwd}F_{+0-0}={1\over\swd} F_{+0+0}\nonumber\\[0.2cm]
  &=&{1\over\cwd} F_{++--}={\ctm s\over16\mwd}\nonumber\\[0.2cm]
      F_{+-+-}&=&-{\ctp\swd s\over16\mwd}\nonumber\\[0.2cm]
      F_{+00-}&=&{\ctp (1-2\cwd)\ s\over32\cw\mwd}\nonumber\\[0.2cm]
      F_{++00}&=&-{ (1-2\cwd)\ s\over16\cw\mwd}\ \ \ \ \ \
      F_{++++}=={\swd s\over8\mwd}
\eqa\\

\vspace{1.5cm}
\centerline{ \fbox{$\g \g\to W^-W^+$}  }
$\fbd$
\bqa
F_{+--+}&=&-{\ctm\cwd s\over32\mwd}\nonumber\ \ \ \ \ \
       F_{+-+-}={\ctp\cwd s\over32\mwd}\nonumber\\[0.2cm]
       F_{+-00}&=&F_{++--}=-{1\over2}F_{++00}={\cwd s\over16\mwd}
\eqa\\

$\fwd$
\bqa
F_{+--+}&=&-{\ctm\swd s\over32\mwd}\nonumber\\[0.2cm]
       F_{+-00}&=&F_{++--}=-{1\over2}F_{++00}=
           {\swd s\over16\mwd}\nonumber\\[0.2cm]
       F_{+-+-}&=&{\ctp\swd s\over32\mwd}
\eqa\\

$\feb$
\bq
  F_{++00}=-{\cw s\over2\mwd}
\eq\\

$\few$
\bq
      F_{++00}=-{\sw s\over2\mwd}
\eq\\

$\fbw$
\bqa
F_{+--+}&=&-{\ctm\cs s\over16\mwd}\ \ \ \ \ \
     F_{+-+-}={\ctp\cs s\over16\mwd} \nonumber\\[0.2cm]
       F_{+-00}&=& F_{++--}=-{1\over2} F_{++00}=
            {\cs s\over8\mwd}
\eqa\\
\newpage
\vspace{1.5cm}
\centerline{ \fbox{$\g W\to\g W$}  }
$\fbd$
\bqa
F_{+-+-}&=&{\ctp\cwd s\over32\mwd}\ \ \ \ \ \
    F_{++++}=-{\cwd s\over16\mwd }    \nonumber\\[0.2cm]
  F_{+0-0}&=&-2 F_{+0+0}=2 F_{++--}=
        {\ctm\cwd s\over16\mwd}
\eqa\\

$\fwd$
\bqa   F_{+-+-}&=&{\ctp\swd s\over32\mwd}\ \ \ \ \ \
   F_{++++}=-{\swd s\over16\mwd}\nonumber\\
       F_{+0-0}&=&=-2 F_{+0+0}=2 F_{++--}
          = {\ctm\swd s\over16\mwd}  \eqa

$\feb$
\bq
       F_{+0-0}={\ctm\cw s\over4\mwd}
\eq\\

$\few$
\bq
 F_{+0-0}={\ctm\sw s\over4\mwd}
\eq\\

$\fbw$
\bqa
F_{+-+-}&=&{\ctp\cs s\over16\mwd}\ \ \ \ \ \
       F_{++++}=-{\cs s\over8\mwd}\nonumber\\[0.2cm]
       F_{+0-0}&=&-2 F_{+0+0}=2 F_{++--}=
            {\ctm\cs s\over8\mwd}
\eqa\\
\newpage

\vspace{1.5cm}
\centerline{ \fbox{$\g \g \to ZZ $}  }
$\fbd$
\bqa
F_{+--+}&=&-{\ctm s\over32\mwd} \ \ \ \ \ \
       F_{+-+-}={\ctp s\over32\mwd}\nonumber\\[0.2cm]
       F_{++--}&=&{ s\over16\mwd}
\eqa\\

$\fwd$
\bqa
F_{+--+}&=&-{\ctm\swd s\over32\cwd\mwd} \ \ \ \ \ \
   F_{+-+-}={\ctp \swd s\over32\cwd\mwd}\nonumber\\[0.2cm]
       F_{++--}&=&{\swd s\over16\cwd\mwd}
\eqa\\

$\fbw$
\bqa
F_{+--+}&=&{\ctm\sw s\over16\cw\mwd} \ \ \ \ \ \
  F_{+-+-}=-{\ctp \sw s\over16\cw\mwd}\nonumber\\[0.2cm]
       F_{++--}&=&-{\sw s\over8\cw\mwd}
\eqa\\
\vspace{1.5cm}
\centerline{ \fbox{$\g Z \to \g Z $}  }
$\fbd$
\bqa
F_{++--}&=&{\ctm s\over32\mwd} \ \ \ \ \ \
 F_{+-+-}={\ctp s\over32\mwd}\nonumber\\[0.2cm]
       F_{++++}&=&-{ s\over16\mwd}
\eqa\\

$\fwd$
\bqa
F_{++--}&=&{\ctm\swd s\over32\cwd\mwd} \ \ \ \ \ \
  F_{+-+-}={\ctp \swd s\over32\cwd\mwd}\nonumber\\[0.2cm]
   F_{++++}&=&-{\swd s\over16\cwd\mwd}
\eqa\\

$\fbw$
\bqa
F_{++--}&=&-{\ctm\sw s\over16\cw\mwd} \ \ \ \ \ \
  F_{+-+-}=-{\ctp \sw s\over16\cw\mwd}\nonumber\\[0.2cm]
       F_{++++}&=&{\sw s\over8\cw\mwd}
\eqa\\

\vspace{1.5cm}
\centerline{ \fbox{$\g Z \to ZZ $} }
$\fbd$
\bqa
F_{+-+-}&=&-F_{+00-}=-{\ctp\sw s\over16\cw\mwd} \ \ \ \ \ \
   F_{++++}=-F_{++00}={ \sw s\over8\cw\mwd}\nonumber\\[0.2cm]
       F_{+--+}&=&F_{+0-0}={\ctm\sw s\over16\cw\mwd}
\eqa\\

$\fwd$
\bqa
F_{+--+}&=&F_{+0-0}=-{\ctm\sw s\over16\cw\mwd} \ \ \ \ \ \
 F_{+-+-}=-F_{+00-}={\ctp\sw s\over16\cw\mwd}\nonumber\\[0.2cm]
        F_{++00}&=&-F_{++++}={\sw s\over8\cw\mwd}
\eqa\\

$\feb$
\bqa
F_{+0-0}&=&-{\ctm s\over16\sw\mwd} \ \ \ \ \ \
 F_{+00-}=-{\ctp s\over16\sw\mwd}\nonumber\\[0.2cm]
        F_{++00}&=&{ s\over8\sw\mwd}
\eqa\\

$\few$
\bqa
F_{+0-0}&=&{\ctm s\over16\cw\mwd} \ \ \ \ \ \
  F_{+00-}={\ctp s\over16\cw\mwd}\nonumber\\[0.2cm]
        F_{++00}&=&-{ s\over8\cw\mwd}
\eqa\\

$\fbw$
\bqa
F_{+--+}&=&F_{+0-0}=-{\ctm (1-2\cwd)\ s\over16\cwd\mwd} \nonumber\\[0.2cm]
 F_{+-+-}&=&-F_{+00-}={\ctp (1-2\cwd)\ s\over16\cwd\mwd}\nonumber\\[0.2cm]
       F_{++00}&=&-F_{++++}={(1-2\cwd)\ s\over8\cwd\mwd}
\eqa\\

\vspace{1.5cm}
\centerline{ \fbox{$ZZ \to ZZ $} }
$\fbd$
\bqa
F_{00++}&=&F_{++00}=-F_{++++}={\swd s\over4\cwd\mwd}\nonumber\\[0.2cm]
F_{0+-0}&=&-F_{+-+-}=F_{+00-}=-{\ctp\swd s\over8\cwd\mwd}\nonumber\\[0.2cm]
  F_{0+0-}&=&F_{+--+}=F_{+0-0}=-{\ctm\swd s\over8\cwd\mwd}
\eqa\\

$\fwd$
\bqa
F_{00++}&=&F_{++00}=-F_{++++}={ s\over4\mwd}\nonumber\\[0.2cm]
F_{0+-0}&=&-F_{+-+-}=F_{+00-}=-{\ctp s\over8\mwd}\nonumber\\[0.2cm]
F_{0+0-}&=&F_{+--+}=F_{+0-0}=-{\ctm s\over8\mwd}
\eqa\\

$\feb$
\bqa
F_{00++}&=&F_{++00}=-{ s\over4\cw\mwd} \ \ \ \ \ \
     F_{0+-0}=F_{+00-}={\ctp s\over8\cw\mwd}\nonumber\\[0.2cm]
  F_{0+0-}&=&F_{+0-0}={\ctm s\over8\cw\mwd}
\eqa\\

$\few$
\bqa
F_{00++}&=&F_{++00}=-{ s\over4\sw\mwd} \ \ \ \ \ \
 F_{0+-0}=F_{+00-}={\ctp s\over8\sw\mwd}\nonumber\\[0.2cm]
  F_{0+0-}&=&F_{+0-0}={\ctm s\over8\sw\mwd}
\eqa\\

$\fbw$
\bqa
F_{00++}&=&F_{++00}=-F_{++++}={\sw s\over2\cw\mwd}\nonumber\\[0.2cm]
F_{0+-0}&=&-F_{+-+-}=F_{+00-}=-{\ctp\sw s\over4\cw\mwd}\nonumber\\[0.2cm]
  F_{0+0-}&=&F_{+--+}=F_{+0-0}=-{\ctm\sw s\over4\cw\mwd}
\eqa\\

\vspace{1.5cm}
\newpage
\centerline{ \fbox{$W^-W^+  \to ZZ $}  }
$\fbd$
\bqa
\cwd F_{0000}&=& F_{++--}= F_{00+-}=-{1\over2}F_{00++}=
{\swd s\over16\mwd}\nonumber\\[0.2cm]
\cw F_{0+-0}&=&- F_{+-+-}=\cw F_{+00-}=-{\ctp\swd s\over32\mwd}
\nonumber\\[0.2cm]
\cw F_{0+0-}&=& F_{+--+}=
\cw F_{+0-0}=-{\ctm\swd\ s\over32\mwd}
\eqa\\

$\fwd$
\bq
F_{0000}={(3-\cos^2\theta)\ s^2\over32 M^4_W}
\eq\\

$\feb$
\bqa
F_{+0-0}&=&F_{0+0-}={\ctm s\over16\mwd} \ \ \ \ \ \
F_{+00-}=F_{0+-0}={\ctp s\over16\mwd}\nonumber\\[0.2cm]
     F_{00++}&=&{ (2\cwd-1)\ s\over4\cw\mwd}
\eqa\\

$\few$
\bqa
F_{++00}&=&{1\over2\cwd-1}F_{00++}=
  -{1\over3}F_{0000}=-{s\over4\sw\mwd}\nonumber\\[0.2cm]
  F_{+0-0}&=&F_{0+0-}=
{\ctm\sw s\over16\cw\mwd}\nonumber\\[0.2cm]
 F_{+00-}&=&F_{0+-0}={\ctp\sw s\over16\cw\mwd}
\eqa\\

$\fbw$
\bqa
F_{++00}&=&-F_{++++}={2\over\swd}F_{00+-}=
-{2\over3}F_{0000}={\sw s\over4\cw\mwd}\nonumber\\[0.2cm]
F_{00++}&=&-2F_{++--}={(1+\cwd)\ \sw s\over4\cw\mwd}\nonumber\\[0.2cm]
F_{0+-0}&=&F_{+00-}={\ctp(2\cwd-1)\ \sw s\over32\cwd\mwd}\nonumber\\[0.2cm]
F_{0+0-}&=&F_{+0-0}={\ctm(2\cwd-1)\ \sw s\over32\cwd\mwd}\nonumber\\[0.2cm]
F_{+--+}&=&-{\ctm s^3_W s\over16\cw\mwd} \ \ \ \ \ \
F_{+-+-}={\ctp s^3_W s\over16\cw\mwd}
\eqa\\

\vspace{1.5cm}
\centerline{ \fbox{$ZW  \to ZW $}  }
$\fbd$
\bqa
F_{0000}&=&{\ctm\swd\ s\over32\cwd\mwd}\nonumber\\[0.2cm]
F_{00++}&=& F_{++00}={ \swd\ s\over16\cw\mwd}\nonumber\\[0.2cm]
F_{+00-}&=&F_{0+-0} =-{ \ctp\swd\ s\over32\cw\mwd}\nonumber\\[0.2cm]
F_{+0-0}&=&-2 F_{+0+0}=2\cw F_{++--}=
 {\ctm \swd s\over16\mwd}\nonumber\\[0.2cm]
F_{-+-+}&=&{\ctp\swd s\over32\mwd} \ \ \ \ \ \
F_{++++}=-{\swd s\over16\mwd}
\eqa\\

$\fwd$
\bq
F_{0000}={(\cos^2\theta-6\ct-3)\ s^2\over64 M^4_W}
\eq\\

$\feb$
\bqa
F_{00++}&=&F_{++00}=-{s\over8\mwd}\ \ \ \ \ \ \
F_{0+-0}=F_{+00-}={\ctp s\over16\mwd} \nonumber\\[0.2cm]
F_{+0-0}&=&-{ (2\cwd-1)\ \ctm s\over8\cw\mwd}
\eqa\\

$\few$
\bqa
F_{0000}&=& 3 F_{0+0-}={3\ctm s\over8\sw\mwd} \ \ \ \ \ \
F_{00++}=F_{++00}=-{\sw s\over8\cw\mwd}\nonumber\\[0.2cm]
F_{+00-}&=&F_{0+-0}={\ctp\sw s\over16\cw\mwd} \ \ \ \ \ \
F_{+0-0}={\ctm (2\cwd-1)\ s\over8\sw\mwd}
\eqa\\

$\fbw$
\bqa
F_{0000}&=&\frac{3}{2}F_{-++-}={3\over\swd}F_{+0+0}={3\over2}F_{0+0-}=
-{3\ctm\sw s\over16\cw\mwd}\nonumber\\[0.2cm]
F_{00++}&=&F_{++00}={(1-2\cwd)\ \sw s\over16\cwd\mwd}\nonumber\\[0.2cm]
F_{0+-0}&=&F_{+00-}={\ctp(2\cwd-1)\ \sw s\over32\cwd\mwd}\nonumber\\[0.2cm]
F_{+0-0}&=&2F_{++--}=-{\ctm(\cwd+1)\ \sw s\over8\cw\mwd}\nonumber\\[0.2cm]
F_{-+-+}&=&{\ctp s^3_W s\over16\cw\mwd} \ \ \ \ \ \
F_{++++}=-{ s^3_W s\over8\cw\mwd}
\eqa\\

\vspace{1.5cm}
\centerline{ \fbox{$W^-W^+  \to W^-W^+$}  }
$\fbd$
\bq
F_{0000}={(3-6\ct-\cos^2\theta)\ s^2\over64M^4_W}
\eq\\

$\fwd$
\bq
F_{0000}={(3-6\ct-\cos^2\theta)\ s^2\over64M^4_W}
\eq\\

$\feb$
\bq
F_{0000}={3(1+\ct)\ s\over8\cw\mwd}
\eq\\

$\few$
\bqa
F_{0000}&=&{3(1+\ct)\ s\over8\sw\mwd} \ \ \ \ \ \
F_{00++}=-{s\over4\sw\mwd}\nonumber\\[0.2cm]
F_{0+0-}&=& F_{+0-0}={\ctm s\over8\sw\mwd}
\eqa\\

$\fbw$
\bqa
F_{0000}&=&F_{00+-}=-F_{0+0+}=F_{+00+}=\nonumber\\[0.2cm]
 &=&F_{0++0}=F_{+-00}=-F_{+0+0}={\ctp\sw\ s\over16\cw\mwd}
\eqa\\

\vspace{1.5cm}
\centerline{ \fbox{$W^+W^+  \to W^+W^+$}  }

$\fbd$
\bq
 F_{0000}={(\cos^2\theta-3)\ s^2\over32M^4_W}
\eq\\

$\fwd$
\bq
  F_{0000}={(\cos^2\theta-3)\ s^2\over32M^4_W}
\eq\\

$\feb$
\bq  F_{0000}=-{3 s\over4\cw\mwd}
\eq\\

$\few$
\bqa
F_{0000}&=&-{3 s\over4\sw\mwd} \ \ \ \ \ \
 F_{+00-}=F_{0+-0}={\ctp s\over8\sw\mwd}\nonumber\\[0.2cm]
  F_{0+0-}&=&F_{+0-0}={\ctm s\over8\sw\mwd}
\eqa\\

$\fbw$
\bqa
F_{0000}&=&F_{00+-}=F_{0++0}=F_{+-00}=\nonumber\\[0.2cm]
 &=&F_{+00+}=-F_{0+0+}=-F_{+0+0}=-{\sw\ s\over8\cw\mwd}
\eqa\\

\subsection{Single Higgs processes}
%
%
%
%
\def\mw{M_W}
\def\mwd{M_W^2}
\def\mwt{M_W^3}
\def\mwq{M_W^4}
\def\mz{M_Z}
\def\mzd{M_Z^2}
\def\ct{\cos\theta\ }
\def\st{\sin\theta\ }
\def\ctd{\cos^2\theta}
\def\ctp{(\cos\theta+1)\ }
\def\ctm{(\cos\theta-1)\ }
\def\cw{c_W}
\def\sw{s_W}
\def\cwd{c_W^2}
\def\swd{s_W^2}
\def\cs{c_W s_W}
\def\rs{\sqrt s}
\def\rd{\sqrt2}
\def\g{\gamma}

\def\fbd{g^2_1\ f^2_B}
\def\fwd{g^2_2\ f^2_W}
\def\feb{e\ g_1 f_B}
\def\few{e\ g_2 f_W}
\def\fbw{g_1 f_B\ g_2 f_W}

The are three helicity indices in the amplitudes now, and the
constrain from parity conservation in the bosonic sector
is given by a relation analogous to (A.1) with the Higgs
treated as longitudinal vector boson.

\vspace{1.5cm}
\centerline{ \fbox{$W^-W^+\to HZ$}  }\noindent
$\fbd$
\bq
F_{000}={\ct\ s^2\over16M_W^4}
\eq\\

$\fwd$
\bq
-F_{\pm 00}=-F_{0\mp 0}={F_{00\mp }\over\cw}
       =\pm {\st\ s\rs\over16\rd\mwt}
\eq\\

$\feb$
\bqa
F_{\pm0\mp}&=&{\ctp s\over16\mwd} \nonumber\\[0.2cm]
F_{0\mp\pm}&=&-{\ctm s\over16\mwd}   \nonumber\\[0.2cm]
F_{000}&=&-{\ct s\over4\cw\mwd}
\eqa\\

$\few$
\bqa
F_{\mp0\pm}&=&{\ctp\sw\ s\over16\cw\mwd}\nonumber\\[0.2cm]
F_{0\mp\pm}&=&-{\ctm\sw\ s\over16\cw\mwd}
\eqa\\

$\fbw$
\bq
-F_{\pm00}=-F_{0\mp0}={F_{00\pm}\over 2\cw}
     =\pm{\st\ \sw\ s\rs\over16\rd\cw\mwt}
\eq\\

\vspace{1.5cm}
\centerline{ \fbox{$ZW \to HW$}  }\noindent
$\fbd$
\bq
F_{000}=-{(\ctd+2\ct-3)\ s^2\over64M_W^4}
\eq\\

$\fwd$
\bq
F_{0\mp0}=\frac{1}{\cw}~ F_{\pm00}=F_{00\pm}=\pm{\st\ s\rs\over16\rd\mwt}
\eq\\

$\feb$
\bqa
F_{\pm\pm0}&=&-{ s\over8\mwd}\nonumber\\[0.2cm]
 F_{\pm 0\mp}&=&-{\ctp s\over16\mwd}\nonumber\\[0.2cm]
    F_{000}&=& {(3+\ct)\ s\over 8\cw\mwd}
\eqa\\

$\few$
\bq
F_{\pm\pm0}=- {\sw\ s\over8\cw\mwd}\ \ \ \ \ \
F_{\pm 0 \mp}=- {\ctp\sw\ s\over16\cw\mwd}
\eq\\

$\fbw$
\bq
F_{0\mp0}={F_{\mp00}\over2\cw}
  =F_{00\pm}=\pm{\st\sw\ s\rs\over16\rd\cw\mwt}
\eq\\

\vspace{1cm}
\centerline{ \fbox{$ H \g \to W^-W^+ $}  }\noindent
$\fbd$
\bq
F_{\pm00}=\mp{\st s_W\ s\rs\over16\rd\mwt}
\eq\\

$\fwd$
\bq
F_{\pm 00}=\pm{\st\sw\ s\rs\over16\rd\mwt}
\eq\\

$\feb$
\bq
F_{\pm\mp0}=-{\ctp\cw\ s\over16\sw\mwd}\ \ \ \ \ \ \ \ \
    F_{\pm0\mp}={\ctm\ c_W s\over16\sw\mwd}
\eq\\

$\few$
\bq
F_{\pm\mp 0}=-{\ctp s\over16\mwd}\ \ \ \ \ \ \ \ \ \ \
    F_{\pm0\mp}={\ctm s\over16\mwd}
\eq\\

$\fbw$
\bq
F_{\pm00}=\pm{\st(1+2\cwd)\ s\rs\over16\rd\cw\mwt}
\eq\\

\vspace{1.5cm}
\centerline{ \fbox{$\g W \to HW $}  }
$\fbd$
\bq
F_{\pm00}=\mp{\st\sw\ s\rs\over16\rd\mwt}
\eq\\

$\fwd$
\bq
F_{\pm00}=\pm{\st\sw\ s\rs\over16\rd\mwt}
\eq\\

$\feb$
\bq
F_{\pm0\mp}={\ctp\cw\ s\over16\sw\mwd}\ \ \ \ \ \ \ \ \ \
    F_{\pm\pm0}={\cw\ s\over8\sw\mwd}
\eq\\

$\few$
\bq
F_{\pm0\mp}={\ctp s\over16\mwd}\ \ \ \ \ \ \ \ \ \ \ \
    F_{\pm\pm0}={ s\over8\mwd}
\eq\\

$\fbw$
\bq
F_{\pm00}=\pm{\st(1+2\cwd)\ s\rs\over16\rd\cw\mwt}
\eq\\

\subsection{Two Higgs processes}
%
%
%

\def\mw{M_W}
\def\mwd{M_W^2}
\def\mwt{M_W^3}
\def\mwq{M_W^4}
\def\mz{M_Z}
\def\mzd{M_Z^2}
\def\ct{\cos\theta}
\def\st{\sin\theta}
\def\ctd{\cos^2\theta}
\def\ctp{(\cos\theta+1)\ }
\def\ctm{(\cos\theta-1)\ }
\def\cw{c_W}
\def\sw{s_W}
\def\cwd{c_W^2}
\def\swd{s_W^2}
\def\cs{c_W s_W}
\def\rs{\sqrt s}
\def\rd{\sqrt2}
\def\g{\gamma}

\def\fbd{g^2_1\ f^2_B}
\def\fwd{g^2_2\ f^2_W}
\def\feb{e\ g_1 f_B}
\def\few{e\ g_2 f_W}
\def\fbw{g_1 f_B\ g_2 f_W}

\vspace{1.5cm}
\centerline{ \fbox{$ HZ \to HZ$} }\noindent
$\fbd$
\bq F_{00}={(\ctd-6\ct-3)\ s^2\over64\mwq}  \eq\\

$\fwd$
\bq F_{00}={(\ctd-6\ct-3)\ s^2\over64\mwq}  \eq\\

$\feb$
\bq F_{+-}={1\over3}F_{00}={\ctm s\over8\cw\mwd} \eq\\

$\few$
\bq F_{+-}={1\over3}F_{00}={\ctm s\over8\sw\mwd} \eq\\

$\fbw$
\bq F_{00}=-2F_{-+}={4\over3}F_{++}=-{\ctm\sw\ s\over4\cw\mwd}
\eq\\

\vspace{1.5cm}
\centerline{ \fbox{$ ZZ \to HH$}}
$\fbd$
\bq F_{00}={(3-\ctd)\ s^2\over32\mwq}  \eq\\

$\fwd$
\bq F_{00}={(3-\ctd)\ s^2\over32\mwq}  \eq\\

$\feb$
\bq F_{--}=-{1\over3}F_{00}=-{ s\over4\cw\mwd} \eq\\

$\few$
\bq F_{--}=-{1\over3}F_{00}=-{ s\over4\sw\mwd} \eq\\

$\fbw$
\bq F_{00}=2F_{++}=-{4\over3}F_{-+}=-{\sw\ s\over2\cw\mwd} \eq\\

\vspace{1.5cm}
\centerline{ \fbox{$ H \g  \to HZ$}}\noindent
$\fbd$
\bq F_{-\mp}=\pm{\ctm\sw\ s\over16\cw\mwd}   \eq\\

$\fwd$
\bq F_{-\pm}=\pm{\ctm\sw\ s\over16\cw\mwd}   \eq\\

$\feb$
\bq F_{-+}=-{\ctm s\over16\sw\mwd}   \eq\\

$\few$
\bq F_{-+}={\ctm s\over16\cw\mwd}   \eq\\
$\fbw$
\bq F_{-\mp}=\pm{\ctm(2\cwd-1)s\over16\cwd\mwd}   \eq\\

\vspace{1.5cm}
\centerline{ \fbox{$ \g Z \to HH $}}
$\fbd$
\bq F_{-\mp}=\pm{\sw\ s\over8\cw\mwd}   \eq\\

$\fwd$
\bq F_{-\pm}=\pm{\sw\ s\over8\cw\mwd}   \eq\\

$\feb$
\bq F_{--}={ s\over8\sw\mwd}   \eq\\

$\few$
\bq F_{--}=-{ s\over8\cw\mwd}   \eq\\

$\fbw$
\bq F_{-\mp}=\pm{(2\cwd-1)\ s\over8\cwd\mwd}   \eq\\

\vspace{1.5cm}
\centerline{ \fbox{$ H\g  \to H\g $}}\noindent
$\fbd$
\bq F_{--}=-{1\over2}F_{-+}=-{\ctm s\over32\mwd}   \eq\\

$\fwd$
\bq F_{--}=-{1\over2}F_{-+}=-{\ctm\swd s\over32\cwd\mwd}   \eq\\

$\fbw$
\bq F_{--}=-{1\over2}F_{-+}={\ctm\sw s\over16\cw\mwd}   \eq\\

\vspace{1.5cm}
\centerline{ \fbox{$ \g \g \to H H$}}
$\fbd$
\bq F_{--}=-2F_{-+}=-{ s\over8\mwd}   \eq\\

$\fwd$
\bq F_{--}=-2F_{-+}=-{\swd s\over8\cwd\mwd}   \eq\\

$\fbw$
\bq F_{--}=-2F_{-+}={\sw s\over4\cw\mwd}   \eq\\
\newpage
\vspace{1cm}
\centerline{ \fbox{$ HW  \to HW $}}\noindent
$\fwd$
\bq F_{00}=-{(3+2\ct- \ctd)\ s^2\over 64\ \mwq}\eq\\

$\few$
\bq F_{-+}={F_{00}\over3}={\ctm\ s\over8\mwd\sw}   \eq\\

\vspace{1.5cm}
\centerline{ \fbox{$ W^- W^+ \to H H$}}\noindent
$\fwd$
\bq F_{00}={(3-\ctd)\ s^2\over32 \mwq}\eq\\

$\few$
\bq F_{++}=-{F_{00}\over3}=-{ s\over4\mwd\sw }
\eq\\
\null

\newpage

\renewcommand{\theequation}{B.\arabic{equation}}
\setcounter{equation}{0}
\setcounter{section}{0}

{\large\bf Appendix B : Helicity amplitudes for Boson-Boson
fusion processes at  High
Energy due to the $\O_{UB}$ interaction}\par
\vspace{1cm}
The non vanishing processes at high ~energies are determined by
the following amplitudes.
\renewcommand{\thesubsection}{B.\arabic{subsection}}
\setcounter{subsection}{0}
\subsection{4-gauge boson processes}
In analogy to the $\O_{UW}$ treatment in \cite{GLRLHC},
it is convenient to express the $\O_{UB}$
contribution to the helicity amplitudes at $s \gsim ~1TeV^2$, as
functions of the initial and final helicities. Below we give the
vector boson fusion amplitudes for the processes
\bq
V_1(\lambda)\  V_2(\lambda') \to V_3(\mu)\  V_4(\mu') \ \ \ ,
 \ \
\eq
where the helicities are indicated in parentheses.
The masses of the vector bosons are denoted by $m_j$  for
$(j=1,...,4)$, while $\epsilon_1$, $\epsilon_2$ denote the
polarization vectors for the initial boson states, and
$\epsilon_3$, $\epsilon_4$ the complex conjugate ones for the
final states. Finally $\theta $ is the c.m.\@ scattering angle;
($z\equiv cos\theta$).
For $s \gsim ~1TeV^2$ we have
\bqa
\ep{1}{2} &=&-\,{s\over2 m_1 m_2}\ (1-\lambda^2)(1-\lambda'^2)
 \nonumber\\
\ep{1}{3} &=&{s(1-\cot)\over4 m_1 m_3}\ (1-\mu^2)(1-\lambda^2)
 \nonumber\\
\ep{1}{4} &=&-\,{s(1+\cot)\over4 m_1 m_4}\ (1-\mu'^2)(1-\lambda^2)
 \nonumber\\
\ep{2}{3} &=&-\,{s(1+\cot)\over4 m_2 m_3}\ (1-\mu^2)(1-\lambda'^2)
 \nonumber\\
\ep{2}{4} &=&{s(1-\cot)\over4 m_2 m_4}\ (1-\mu'^2)(1-\lambda'^2)
\nonumber\\
\ep{3}{4} &=&-\,{s\over2 m_3 m_4}\ (1-\mu^2)(1-\mu'^2)
 \ \ , \ \ \ \ \ \eqa
and the definitions
\bqa
V_{12}&=&{s\over4}\, \lambda^2\lambda'^2(1+\lambda\lambda')\ \ \ \ ;
\ \ \
 V_{34}={s\over4}\, \mu^2\mu'^2(1+\mu\mu')\nonumber\\
V_{13}&=&{s\over8}\,(1-\lambda\mu)(1-\cot)\mu^2\lambda^2\ \ \ \ ;\ \ \
V_{24}={s\over8}\,(1-\lambda'\mu')(1-\cot)\mu'^2\lambda'^2 \nonumber\\
V_{14}&=&{s\over8}\,(1-\lambda\mu')(1+\cot)\mu'^2\lambda^2\ \ \ \ ;
\ \ \
V_{23}={s\over8}\,(1-\lambda'\mu)(1+\cot)\mu^2\lambda'^2
\eqa
\bq
Z^B_{ij}=\ep{i}{j}\ {M_W\over\cwd} - {2d_B\over M_W}\swd\ V_{ij}
\eq
The Higgs propagator is written as
\bq D_H(x)\equiv x-{\cal M}^2_H  \eq
where ${\cal M}^2_H\equiv M^2_H$ for $x=t$ or $u$, and ${\cal
M}^2_H\equiv M^2_H-i\ M_H\Gamma_H$ when $x=s$. \par

We find
\bq
F_H(\gamma W\to\gamma W)\ =\ 2g^2_2 c^2_W d_B
V_{13}\ep{2}{4}
 \dh{t} \eq
\bq F_H(\gamma W\to ZW)\ =\ -{s_W\over c_W}\, F_H(\gamma W\to\gamma W)
 \eq
\bq F_H(Z W\to \gamma W)\ =\ F_H(\gamma W\to Z W) \eq

\bq   F_H(ZW\to ZW)=-g^2_2 Z^B_{13}\ep{2}{4}\ M_W\dh{t} \eq
\bq F_H(\gamma \gamma\to WW)= 2g^2_2 c^2_W d_BV_{12}\ep{3}{4}
\dh{s}  \eq
\bq F_H(\gamma Z\to WW)=-{s_W\over c_W} F_H(\gamma\gamma\to WW) \eq

\bq  F_H(ZZ\to WW)=-g^2_2 Z^B_{12}\ep{3}{4}M_W\dh{s} \eq

\bqa  F_H(ZZ\to ZZ)&=&-g^2_2\biggm\{
Z^B_{12}Z^B_{34}\dh{s}+ \nonumber\\
& & Z^B_{13}Z^B_{24}\dh{t}
  +Z^B_{14}Z^B_{23}\dh{u} \biggm\} \eqa

\bq
F_H(W^+W^-\to\gamma\gamma)=2g^2_2 c^2_W d_B V_{34}\ep{1}{2}
\dh{s}  \eq

\bq
F_H(W^+W^-\to\gamma Z)=-{s_W\over c_W}
F_H(W^+W^-\to\gamma\gamma) \eq

\bq
   F_H(W^+W^-\to
ZZ)=-g^2_2\ep{1}{2}Z^B_{34}M_W
\dh{s}\eq

\bq F_H(\gamma\gamma\to\gamma\gamma)=
-{4g^2_2c^4_Wd_B^2\over M^2_W}\bigg\{ {V_{12}V_{34}\over\nh{s} }
 +{V_{13}V_{24}\over\nh{t}}+{V_{14}V_{23}\over\nh{u} } \biggm\}\eq

\bq F_H(\gamma Z\to\gamma\gamma)=-{s_W\over c_W}
F_H(\gamma\gamma\to\gamma\gamma) \eq

\bq F_H(ZZ\to\gamma\gamma)=
g^2_2\bigg\{{2c^2_W d_B\over M_W}\ {Z^B_{12}V_{34}\over\nh{s} }
 -{4 s^2_W c^2_W d^2_B\over
M^2_W}\co{V_{13}V_{24}\over\nh{t}}+{V_{14}V_{23}\over
\nh{u}}\cf
 \biggm\}\eq

\bq F_H(\gamma \gamma\to\gamma Z)=-{s_W\over c_W}
F_H(\gamma\gamma\to\gamma\gamma) \eq

\bq F_H(\gamma Z\to\gamma Z)=
g^2_2\bigg\{{2c^2_W d_B\over M_W} {Z^B_{24}V_{13}\over\nh{t} }
 -{4 c^2_W s^2_W d^2_B\over M^2_W}\co{V_{12}V_{34}\over\nh{s}}+{V_{14}V_{23}
\over\nh{u}}\cf
 \biggm\}\eq

\bq F_H(Z Z\to\gamma Z)=
-2g^2_2{s_Wc_Wd_B\over M_W}\,\bigg\{ {Z^B_{12}V_{34}\over\nh{s} }
 +{V_{13}Z^B_{24}\over\nh{t}}+{Z^B_{14}V_{23}\over\nh{u} } \biggm\}\eq

\bq F_H(\gamma \gamma\to Z Z)=
{2g^2_2c^2_W d_B\over M_W}\,\bigg\{ {Z^B_{34}V_{12}\over\nh{s} }
 -{2 s^2_W d_B\over M_W}\co{V_{13}V_{24}\over\nh{t}}+{V_{14}V_{23}
\over\nh{u}}\cf
 \biggm\}\eq

\bq F_H(\gamma Z\to Z Z)=
-{2g^2_2 s_Wc_W d_B\over M_W}\,\bigg\{ {V_{12}Z^B_{34}\over\nh{s} }
 +{V_{13}Z^B_{24}\over\nh{t}}+{V_{14}Z^B_{23}\over\nh{u} } \biggm\}\eq

\subsection {Higgs production processes}
No single Higgs process due to $\O_{UB}$ survives at high energy. We have
only to consider two Higgs processes.\par

     First the  processes $V_1(\lambda)V_2(\tau) \to HH$ are described by
9 helicity amplitudes  $F_{\lambda \tau}(\theta)$. Here
$\theta$ is the angle between $V_1$ and H, and the normalization is
such that the differential cross section writes\\
\bq     {d\sigma(\lambda \tau )\over d cos(\theta)} =
 C|F_{\lambda \tau}(\theta)|^2  \ \ \ \ \ , \ \ \ \eq\\
where\\
\bq  C = {1\over32\pi s}\,{p_{H}\over p_{12}} \ \ \ \ \ , \ \ \
 \eq\\
includes \underline{no spin average}. We find\par
\vspace{1.0cm}

\fbox{ {\bf $ ZZ, \gamma\gamma, \gamma Z \to HH $ }}
\bqa
 F_{\lambda \tau}(\theta)&=& -(1-\delta_{\tau 0})
(1-\delta_{\lambda 0})( s^2_W, c^2_W,
-\sw \cw) . \nonumber \\[0.5cm]
  & & \biggm\{{d^2_B g^2_2 s\over 2\mwd
}(1+3\lambda\tau)+{d_B g^2_2s\over
4\mwd}(1+\lambda\tau)\biggm\}
\eqa
\vspace{1cm}
and for the crossed channel\\
\fbox{ {\bf \, $HZ \to HZ$,\,$ H\gamma \to H\gamma$,
\,$ H\gamma \to HZ $ }}\par
\vspace{0.5cm}
These $H V_1(\tau) \to HV_2(\mu)$ channels are obtained by
crossing those above. The helicity amplitudes are now
given by\\
\bqa
F_{\tau \mu}(\theta) &=& -(1-\delta_{\mu 0})
(1-\delta_{\tau 0})(s^2_W, c^2_W,
-\sw \cw)(1-\cot) . \nonumber \\[0.5cm]
  & &\biggm\{{d^2_B g^2_2 s\over
4\mwd}(1-3\tau\mu)+{d_B g^2_2s\over
8\mwd}(1+\tau\mu)\biggm\}
\eqa

\end{document}